\documentclass[11pt,a4paper]{article}

\usepackage{epsfig,amsmath,amssymb,cite}
\usepackage{hyperref}

\begin{document}

% ORCID Griselda Figueroa-Aguirre: https://orcid.org/0000-0002-3477-2477

\title{Thin--shell wormholes in $N$--dimensional $F(R)$ gravity } 
\author{ Griselda Figueroa-Aguirre$^{1}$\thanks{e-mail: gfigueroa@iafe.uba.ar}\\
{\small $^1$ Instituto de Astronom\'{\i}a y F\'{\i}sica del Espacio (IAFE, CONICET-UBA),}\\
{\small Ciudad Universitaria, 1428 Buenos Aires, Argentina}}
\date{}
\maketitle

\begin{abstract}

In this work, spherically symmetric thin--shell wormholes with a conformally invariant Maxwell field for $N$--dimensional $F(R)$ gravity and constant scalar curvature $R$ are built. Two cases are considered: wormholes symmetric across the throat and asymmetric ones having different values of the scalar curvature across the throat. Their stability under radial perturbations is analyzed, finding that unstable and stable solutions are possible for suitable values of the parameters, always made of exotic matter. The stable solutions are found for a short range, slightly over a large critical value of charge.

\end{abstract}


\section{Introduction}\label{intro}

By adding terms to the gravitational Lagrangian, modified gravity theories have been proposed as alternatives to explain cosmological features such as the accelerated expansion of the universe or the early time inflation without resorting to exotic elements such as dark energy. Due to their relative simplicity, one of the most popular among them are the F(R) theories \cite{sofa1, sofa2,sofa3,sofa4}, which introduce a modification in the general relativity action by replacing the Ricci scalar $R$ by a function $F(R)$. Many solutions in four dimensions have been found in this theory: among them, traversable wormholes \cite{whfr1_1, whfr1_2, whfr2a, whfr2b, whfr3, bhfr3_8} and spherically symmetric black holes \cite{bhfr1a, bhfr2b,bhfr2a, bhfr1b, bhfr2c, bhfr2d, bhfr2e, bhfr3_1, bhfr3_2, bhfr3_3, bhfr3_4, bhfr3_5, bhfr3_6, bhfr3_7,bhfr3_8}. Additional solutions have been explored in higher dimensions \cite{BHFRndim1a, BHFRndim1b,BHFRndim2,BHFRndim3,tsndim} as well as in lower ones\cite{BHFR3dim}. Some of these solutions can be reduced to the black hole solutions in general relativity \cite{BHRGndim1, BHRGndim2, BHRGndim3, BHRGndim4, BHRG3dim1, BHRG3dim2, BTZ92} within the adequate limits. 

In the previous century, the Darmois--Israel formalism \cite{darisa, darisb} has been developed as a technique which allows to craft a new manifold by cutting two spacetimes and pasting them onto a hypersuface made of a thin layer of matter. In order to obtain a proper matching, certain particular conditions need to be fulfilled. Via the analysis of the energy--momentum tensor at this hypersurface, its stability can be analytically studied when it is done under perturbations that preserve the symmetry. Due to the popularity of this formalism, several applications of it can be found in the literature: for four--dimensional objects, one can list gravastars \cite{gravstar1, gravstar2, gravstar3}, bubbles \cite{bubbles}, wormholes \cite{whrg1, whrg2, whrg3, whrg4, whrg5, whrg6} and thin shells enclosing black holes \cite{shrg1, shrg2, shrg3, shrg5, shrg4}, while for higher dimensions, thin shells of matter and wormholes  \cite{GRddim1, GRddim2, GRddim3, GRddim4}. 

Over the years, this formalism has been extended to other theories;  in particular for metric $F(R)$ gravity, a general set of junction conditions has been developed \cite{dss,js1, js2}, showing that they have a more restrictive nature than those in general relativity. They always demand the continuity of the trace of the second fundamental form at the joining hypersurface. Depending on the third derivative of $F(R)$, the formalism branches out: for  $F'''(R) \neq 0$ the continuity of the scalar curvature $R$ on the hypersurface is also demanded; if $F'''(R) = 0$, it is possible to construct spacetimes with a discontinuity of the scalar curvature across the hypersurface, in which case, extra contributions to the energy--momentum tensor appear in order to guarantee its local conservation \cite{js1,js2}. Applications of this formalism in $F(R)$ gravity can be found in the literature; namely, thin shells and wormholes in four dimensions \cite{TSWHinFR1, TSWHinFR2, TSWHinFR3, TSWHinFR4, TSWHinFR5, tsFR1, tsFR2, tsFR3, god1, god2, god3, god5} as well as in lower dimensions \cite{tsFRlowdim1, tsFRlowdim2,tsFRlowdim3}. There are also works in non-local theories of gravity \cite{god4} as well as in $F(R,T)$ \cite{god6}. Thin shells in higher dimensional $F(R)$ gravity have been analyzed \cite{tsndim}, but there are no analogous studies with wormholes; however, it has been done for General Relativity \cite{GRddim1}. The main motivation to work with higher dimensional spacetimes is provided by string theory, which proposes a path to quantum gravity and to a unified theory of physics \cite{mot74a}. The AdS/CFT correspondence relates string theory in a $N$-dimensional asymptotically anti-de Sitter background with a conformal quantum field theory on the boundary having $N-1$ dimensions \cite{mot74b}. Cosmological models with extra dimensions have been introduced in order to provide possible explanations to open problems \cite{mot74c}. The study of higher dimensional black holes \cite{mot75} contributes to a better understanding of these objects, because gravity is richer in more than four dimensions. Wormhole spacetimes in higher dimensions have been analyzed in General Relativity (e.g. Ref. \cite{mot76a} and references therein), in modified theories of gravity (e.g. \cite{mot76b} references therein), and within the framework of braneworld cosmology \cite{mot76c}. 

In this work, spherically symmetric thin--shell wormholes with a conformally invariant Maxwell field for $N$--dimensional metric $F(R)$ gravity are constructed. In Sects. \ref{construction} and \ref{stability}, the general formalism and the procedure for the analysis of stability of static configurations under perturbations that preserve the symmetry are described for a large family of geometries. The generalized charged black hole solution \cite{tsndim} with constant $R$, that is used for the construction of the wormholes, is introduced in Sect. \ref{sol-fr}.  Along Sect. \ref{examples}, two examples of $N$--dimensional thin--shell wormholes are developed: symmetric across the shell and asymmetric ones in the value of $R$. In Sect. \ref{summary}, a summary of the work is presented.
Along this article, the proper time derivative is denoted with a dot over a given quantity $\Upsilon$, that is, $d \Upsilon /d\tau = \dot{\Upsilon}$, while the radial derivative of the metric function $A(r)$ is assigned with a prime, that is, $d A(r)/dr= A'(r) $. However, the prime on the $F(R)$ describes the derivative with respect to the scalar curvature $R$, that is, $d F(R)/dR=F'(R)$. Units are taken such as $c=G_N=1$, and the metric signature is $(-,+,\dots,+)$.

\section{Wormhole construction} \label{construction}

In order to construct a spherically symmetric wormhole in $N$--dimensional $F(R)$ gravity, two metrics with the following form are adopted 
\begin{equation}
ds^2 = -A_{1,2}(r)dt_{1,2}^2 + A_{1,2}(r)^{-1}dr^2 +r^2d\Omega^{N-2}, \quad
d\Omega^2_{N-2}= d\theta^2_1+\sideset{}{}\sum_{i=2}^{N-2} \sideset{}{ }\prod_{j=1}^{i-1} \sin^2 \theta _j d\theta_i^2,
\label{metric-N-sphe}
\end{equation}
where $t_{1,2}$ are the respecting time coordinates, $r>0$ is the radial coordinate, $0\le \theta _i\le \pi$ ($1\le i \le N-3$) and $0\le \theta _{N-2} < 2 \pi $ are the angular coordinates. 

Two manifolds $\mathcal{M}_1$ and $\mathcal{M}_2$, with constant scalar curvature $R_{1,2}$ and described by Eq. (\ref{metric-N-sphe}), are defined by cutting at the same radius $a$ and taking $r\geq a$ (with $a$ large enough in order to remove horizons and singularities). They are pasted together onto a hypersurface $\Sigma$ defined by $G(r)=r-a=0$. The new manifold $\mathcal{M}=\mathcal{M}_1 \cup \mathcal{M}_2$ is the union of them, with values $R_{1,2}$ at each side of $\Sigma$. Since these spaces are spherically symmetric, and the area of a $(N-2)$--sphere is proportional to $r^{N-2}$, it is always minimal for $r=a$ so the flare--out condition is fulfilled, guaranteeing that the construction describes a wormhole with its throat at $a$. 
The coordinates of the original manifolds $\mathcal{M}_{1,2}$ are  $X^{\alpha }_{1,2} = (t_{1,2},r,\theta_1, ... ,\theta_{N-2})$, and due to the spherical symmetry, the angular coordinates of each manifold have been mutually identified. A new global radial coordinate $\ell$ in $\mathcal{M}$ can be defined, that is, $\ell=\pm \int_a^{r}\sqrt{1/A_{1,2}(r)} dr$, where $(+)$ sign corresponds to $\mathcal{M}_{1}$ and $(-)$ to $\mathcal{M}_{2}$, and $\ell=0$ is the position of the throat. On the hypersurface $\Sigma$, the adopted coordinates are $\xi ^{i}=(\tau ,\theta_1, ... ,\theta_{N-2})$, with $\tau $ the proper time there. The throat radius $a$ is assumed to be a function of $\tau $ in order to study the stability of the static configurations. The proper time $\tau $ can be related to the coordinate times of the two manifolds through the expression
\begin{equation}
\frac{dt_{1,2}}{d\tau} = \frac{\sqrt{A_{1,2}(a) + \dot{a} ^2}}{A_{1,2}(a)},
\label{tau}
\end{equation}
where the signs are determined by choosing $t_{1,2}$ and $\tau$ to run into the direction of the future. 

The first fundamental form for the original manifolds can be calculated as 
\begin{equation}
h^{1,2}_{ij}= \left. g^{1,2}_{\mu\nu}\frac{\partial X^{\mu}_{1,2}}{\partial\xi^{i}}\frac{\partial X^{\nu}_{1,2}}{\partial\xi^{j}}\right| _{\Sigma },
\end{equation}
the extrinsic curvature or second fundamental form by
\begin{equation}
K_{ij}^{1,2 }=-n_{\gamma }^{1,2 }\left. \left( \frac{\partial ^{2}X^{\gamma
}_{1,2} } {\partial \xi ^{i}\partial \xi ^{j}}+\Gamma _{\alpha \beta }^{\gamma }
\frac{ \partial X^{\alpha }_{1,2}}{\partial \xi ^{i}}\frac{\partial X^{\beta }_{1,2}}{
\partial \xi ^{j}}\right) \right| _{\Sigma },
\label{sff}
\end{equation}
where the unit normals ($n^{\gamma }n_{\gamma }=1$) are obtained via
\begin{equation}
n_{\gamma }^{1,2 }=\left\{ \left. \left| g^{\alpha \beta }_{1,2}\frac{\partial G}{\partial
X^{\alpha }_{1,2}}\frac{\partial G}{\partial X^{\beta }_{1,2}}\right| ^{-1/2}
\frac{\partial G}{\partial X^{\gamma }_{1,2}} \right\} \right| _{\Sigma }.
\end{equation}
For an easy interpretation of the physical magnitudes, this  construction is described by using the following orthonormal basis 
\begin{equation*}
e_{\hat{\tau}}=e_{\tau }, \qquad e_{\hat{\theta}_1}=a^{-1}e_{\theta_1 }, \qquad e_{\hat{\theta}_i}=\left( a\prod_{j=1}^{i-1} \sin \theta _j \right)^{-1} e_{\theta_i } \quad \mathrm{if} \quad 2\le i \le N-2 .
\end{equation*}
When all these elements are calculated by using the metrics given by Eq. (\ref{metric-N-sphe}), the first fundamental form becomes 
\begin{equation}
h^{1,2}_{\hat{\imath}\hat{\jmath}}= \mathrm{diag}(-1,1,...,1), 
\label{h-metric}
\end{equation}
the unit normals read 
\begin{equation} 
n_{\gamma }^{1,2}=  \pm \left(-\dot{a},\frac{\sqrt{A_{1,2}(a)+\dot{a}^2}}{A_{1,2}(a)},0,...,0 \right),
\end{equation}
and the non--null components of the extrinsic curvature are given by
\begin{equation} 
K_{\hat{\tau}\hat{\tau}}^{1,2 }=  \mp \frac{A '_{1,2}(a)+2\ddot{a}}{2\sqrt{A_{1,2}(a)+\dot{a}^2}},
\label{e5}
\end{equation}
and
\begin{equation} 
K_{\hat{\theta}_i\hat{\theta}_i}^{1,2}= \pm\frac{1}{a}\sqrt{A_{1,2} (a) +\dot{a}^2}.
\label{e4}
\end{equation}

The junction formalism in metric $F(R)$ gravity \cite{js1} requires the fulfillment of several conditions in order to obtain a proper matching hypersurface, namely 
\begin{itemize}
\item The first fundamental form must be continuous at  $\Sigma $, or in other words, the jump\footnote{Expressions within brackets such as $[\Upsilon ]\equiv (\Upsilon ^{2}-\Upsilon  ^{1})|_\Sigma $ are used to define the jump of any quantity $\Upsilon $ across the hypersurface $\Sigma$. } of $h_{\mu \nu}$ has to be zero,
\begin{equation}
[h_{\mu \nu}]=0.
\label{1rstcondition}
\end{equation}
This condition is automatically satisfied in the current construction thanks to Eq. (\ref{h-metric}).
\item The trace of the second fundamental form should be continuous at $\Sigma $, 
\begin{equation}
[K^{\mu}_{\;\; \mu}]=0.
\label{2rstcondition}
\end{equation}
With the use of Eqs. (\ref{e5}) and (\ref{e4}), this condition can be translated to
\begin{equation} 
\frac{2\ddot{a}+ A_{2}'(a)}{2\sqrt{A_{2}(a)+\dot{a}^2}}+\frac{2\ddot{a}+ A_{1}'(a)}{2\sqrt{A_{1}(a)+\dot{a}^2}}+\frac{(N-2)}{a}\left(\sqrt{A_{2}(a)+\dot{a}^2}+\sqrt{A_{1}(a)+\dot{a}^2}\right)=0.
\label{CondGen}
\end{equation}
\item A third condition may be required, depending on the value of the third derivative of $F(R)$. Here is where the formalism develops into two branches: when  $F'''(R) \neq 0$ \cite{js1,js2}, the construction requires 
\begin{equation}
[R]=0.
\label{3rstcondition}
\end{equation}
However, this condition can be ignored if $F'''(R)=0$. The way in which the formalism  for the construction of a thin--shell wormhole branches out is explained in the following subsections. 
\end{itemize}

\subsection{Case $F'''(R) \neq 0$}

When  $F'''(R) \neq 0$, Eq. (\ref{3rstcondition}) should be satisfied and the field equations \cite{js1} at $\Sigma$ take the form
\begin{equation} 
\kappa S_{\mu \nu}=-F'(R)[K_{\mu \nu}]+ F''(R)[\eta^\gamma \nabla_\gamma R]  h_{\mu \nu}, \;\;\;\; n^{\mu}S_{\mu\nu}=0,
\label{LancGen}
\end{equation}
with $\kappa =8\pi $ and $S_{\mu \nu}$ the energy--momentum tensor at $\Sigma$. When the scalar curvature is constant, this expression simplifies to
\begin{equation} 
\kappa S_{\mu \nu}=-F'(R_0)[K_{\mu \nu}], \;\;\;\; n^{\mu}S_{\mu\nu}=0.
\label{LanczosGen}
\end{equation}
The energy--momentum tensor, in the orthonormal basis, takes the diagonal form $S_{_{\hat{\imath}\hat{\jmath} }}=\mathrm{diag}(\sigma ,p,...,p)$ , with $\sigma$ the hypersurface energy density and $p \equiv p_{\hat{\theta}_i}$ ($1 \le i \le N-2$) the transverse pressure. Their expressions read
\begin{equation} 
\sigma= \frac{F'(R_0)}{\kappa } \left( \frac{2\ddot{a}+A_{2}'(a)}{2\sqrt{A_{2}(a)+\dot{a}^2}} + \frac{2\ddot{a}+A_{1}'(a)}{2\sqrt{A_{1}(a)+\dot{a}^2}}\right),
\label{e9}
\end{equation}
and
\begin{equation}
p=\frac{- F'(R_0)}{\kappa }\left(\frac{\sqrt{A_{2}(a)+\dot{a}^2}}{a}+\frac{\sqrt{A_{1}(a)+\dot{a}^2}}{a}\right).
\label{e10}
\end{equation}
With the help of Eqs. (\ref{e9}) and (\ref{e10}), Eq. (\ref{CondGen}) can be rewritten, resulting  after some algebra, in the equation of state $\sigma -(N-2)p=0$.

\subsection{Case $F'''(R) = 0$}

Due to $F'''(R) = 0$, it is easy to see that the construction is made within a quadratic $F(R)$ theory, that is, $F(R)=R-2\Lambda+\gamma R^2$ and, therefore, $F'(R)= 1+2\gamma R$. For this particular case, the continuity of the scalar curvature given by the Eq. (\ref{3rstcondition}) is not necessary \cite{js1,js2}. The field equations \cite{js1} at $\Sigma$ are given by
\begin{equation}
\kappa S_{\mu \nu} =-[K_{\mu\nu}]+2\gamma \left( [n^{\gamma }\nabla_{\gamma}R] h_{\mu\nu}-[RK_{\mu\nu}] \right), \;\;\;\; n^{\mu}S_{\mu\nu}=0.
\label{LancQuad}
\end{equation}
However, when the scalar curvature $R_{1,2}$ is constant at $\Sigma$, this expression can be simplified to
\begin{equation}
\kappa S_{\mu \nu} =-[K_{\mu\nu}]-2\gamma[RK_{\mu\nu}], \;\;\;\; n^{\mu}S_{\mu\nu}=0.
\label{LanczosQuad}
\end{equation}
In order to guarantee a divergence--free energy--momentum tensor, and therefore, local conservation, extra contributions are necessary. The energy--momentum tensor at $\Sigma$ becomes $(S_{\mu \nu} + \mathcal{T} n_\mu n_\nu + \mathcal{T}_\mu n_\nu + \mathcal{T}_\nu n_\mu )\delta ^{\Sigma } + \mathcal{T}_{\mu \nu}$, with $\delta ^{\Sigma }$ the Dirac delta on $\Sigma $, and
\begin{itemize}
\item $\mathcal{T}$ an external scalar pressure or tension
\begin{equation}
\kappa\mathcal{T}=2\gamma [R] K^\gamma{}_\gamma ,
\label{Tg}
\end{equation}
\item $\mathcal{T}_\mu$ an external energy flux vector,
\begin{equation}
\kappa\mathcal{T}_\mu=-2\gamma \bar{\nabla}_\mu[R]=0,  \qquad  n^{\mu}\mathcal{T}_\mu=0,
\label{Tmu}
\end{equation}
where $\bar{\nabla }$ the intrinsic covariant derivative on $\Sigma$.  This element is zero when both $R_{1,2}$ are constant. 
\item $\mathcal{T}_{\mu \nu}$ a two--covariant symmetric tensor distribution
\begin{equation}
\kappa \mathcal{T}_{\mu \nu}=\nabla_{\beta } \left( 2\gamma [R] h_{\mu \nu } n^{\beta } \delta ^{\Sigma }\right),
\label{dlay1}
\end{equation}
which has a resemblance with classical dipole distributions.
\end{itemize}
For more details about these additional contributions, Refs. \cite{js1,js2} are recommended.

Once again, due to the use of the orthonormal basis, the energy--momentum tensor $S_{_{\hat{\imath}\hat{\jmath} }}=\mathrm{diag}(\sigma ,p,...,p)$ has a diagonal form, allowing an easy identification with the physical magnitudes of the energy density $\sigma$ and the pressure $p$ on the hypersurface, 
\begin{equation} 
\sigma= \frac{1+2\gamma R_2}{\kappa }\left( \frac{2\ddot{a}+A_{2}'(a)}{2\sqrt{A_{2}(a)+\dot{a}^2}} \right)+ \frac{1+2\gamma R_1}{\kappa }\left( \frac{2\ddot{a}+A_{1}'(a)}{2\sqrt{A_{1}(a)+\dot{a}^2}}\right),
\label{e9Rdif}
\end{equation}
and
\begin{equation}
p= -\frac{1+2\gamma R_2}{\kappa }\left(\frac{\sqrt{A_{2}(a)+\dot{a}^2}}{a}\right)- \frac{1+2\gamma R_1}{\kappa }\left(\frac{\sqrt{A_{1}(a)+\dot{a}^2}}{a}\right).
\label{e10Rdif}
\end{equation}
With Eq. (\ref{CondGen}) and (\ref{Tg}), a symmetrical form of the external scalar pressure or tension $\mathcal{T} $ can be written 
\begin{eqnarray}
\mathcal{T} &=& \frac{2\gamma R_2}{\kappa }\left( \frac{2\ddot{a}+A_{2}'(a)}{2\sqrt{A_{2}(a)+\dot{a}^2}} + (N-2) \frac{\sqrt{A_{2}(a)+\dot{a}^2}}{a} \right) \nonumber \\
& & + \frac{2\gamma R_1}{\kappa }\left( \frac{2\ddot{a}+A_{1}'(a)}{2\sqrt{A_{1}(a)+\dot{a}^2}}+ (N-2)\frac{\sqrt{A_{1}(a)+\dot{a}^2}}{a}\right).
\label{e11Rdif}
\end{eqnarray} 
The tensor distribution $\mathcal{T}_{\hat{\imath}\hat{\jmath}}$ is proportional to $2 \gamma [R] h_{\hat{\imath}\hat{\jmath}} /\kappa$. Using Eqs. (\ref{e9Rdif}), (\ref{e10Rdif}), and (\ref{e11Rdif}), the expression of Eq. (\ref{CondGen}) can be rewritten, obtaining the equation of state $\sigma - (N-2) p=\mathcal{T}$ for this case. 
The matter at the throat of the thin--shell wormhole can be characterized by using the weak energy condition (WEC); as long as it satisfies $\sigma \geq 0$ and $\sigma + p \geq 0$, it is normal; otherwise it is exotic. Along this formalism, another requirement has been taken into consideration: $F'(R)>0$, which is necessary within $F(R)$ gravity to guarantee a positive effective Newton constant $G_{eff} = G/F'(R_0)$, and therefore, to avoid the presence of ghosts \cite{bronnikov}.

\section{Stability analysis} \label{stability}

For the stability analysis, radial perturbations of the static radius $a_0$ of the throat are considered.
In all cases, the condition of the continuity of the trace of the second fundamental form, given by Eq. (\ref{CondGen}), reduces to 
\begin{equation} 
\frac{ A_{2}'(a_0)}{2\sqrt{A_{2}(a_0)}}+\frac{ A_{1}'(a_0)}{2\sqrt{A_{1}(a_0)}}+\frac{(N-2)}{a_0}\left(\sqrt{A_{2}(a_0)}+\sqrt{A_{1}(a_0)}\right)=0.
\label{CondEstatico}
\end{equation}
When $F'''(R) \neq 0$, the static values of the energy density $\sigma _0$ and the pressure $p_0$ are
\begin{equation} 
\sigma_0= \frac{F'(R_0)}{\kappa }\left( \frac{A_{2}'(a_0)}{2\sqrt{A_{2}(a_0)}}+\frac{A_{1}'(a_0)}{2\sqrt{A_{1}(a_0)}}\right),
\label{e13}
\end{equation}
and
\begin{equation}
p_0=-\frac{- F'(R_0)}{a_0\kappa }\left( \sqrt{A_{2}(a_0)}+\sqrt{A_{1}(a_0)}\right),
\label{e14}
\end{equation}
which satisfy the equation of state $\sigma_0 - (N-2) p_0=0$. 

For $F'''(R) = 0$, the energy density $\sigma _0$, the pressure $p_0$, and the external tension/pressure $\mathcal{T}_0$ take the form
\begin{equation} 
\sigma_0= \frac{1+2\gamma R_2}{\kappa }\left( \frac{A_{2}'(a_0)}{2\sqrt{A_{2}(a_0)}} \right)+ \frac{1+2\gamma R_1}{\kappa }\left( \frac{A_{1}'(a_0)}{2\sqrt{A_{1}(a_0)}}\right),
\label{e13DifR}
\end{equation}
\begin{equation}
p_0= -\frac{1+2\gamma R_2}{\kappa }\left(\frac{\sqrt{A_{2}(a_0)}}{a_0}\right)- \frac{1+2\gamma R_1}{\kappa }\left(\frac{\sqrt{A_{1}(a_0)}}{a_0}\right),
\label{e14DifR}
\end{equation}
and
\begin{equation}
\mathcal{T}_0=\frac{2\gamma R_2}{\kappa }\left( \frac{A_{2}'(a_0)}{2\sqrt{A_{2}(a_0)}} + (N-2) \frac{\sqrt{A_{2}(a_0)}}{a_0} \right)+ \frac{2\gamma R_1}{\kappa }\left( \frac{A_{1}'(a_0)}{2\sqrt{A_{1}(a_0)}}+ (N-2)\frac{\sqrt{A_{1}(a_0)}}{a_0}\right).
\label{e16DifR}
\end{equation}
The equation of state now reads $\sigma_0 - (N-2) p_0=\mathcal{T}_0 $. The other non--null extra contribution is given by the double layer tensor distribution $\mathcal{T}^{(0)}_{\hat{\imath}\hat{\jmath}}$, which is proportional to $2 \gamma (R_2-R_1) h_{\hat{\imath}\hat{\jmath}} /\kappa$. In this case, a double layer of matter at the throat coexists with the thin shell.

To analyze the stability of these constructions under radial perturbations, it is useful to obtain an effective potential $V(a)$  related to $\dot{a}^2$ in the following form
\begin{equation}
\dot{a}^{2}=-V(a).
\label{condicionPot}
\end{equation}
The expression of Eq. (\ref{CondGen}) can be rewritten by considering  $\ddot{a}= (1/2)d(\dot{a}^2)/da$ and by using the definition $z=\sqrt{A_{2}(a)+\dot{a}^2}-\sqrt{A_{1}(a)+\dot{a}^2}$, obtaining 
\begin{equation}
az'(a)+2z(a)=0.
\label{CondGen_u}
\end{equation}
Solving the differential equation above results in an expression for $\dot{a}^{2}$ in terms of $a$ and, therefore, for the potential in Eq. (\ref{condicionPot}), which has the form
\begin{equation}
V(a)= -\frac{a_{0}^{2N-4}\left(\sqrt{A_{2}(a_{0})}+\sqrt{A_{1}(a_{0})}\right)^2}{4a^{2N-4}} +\frac{A_{1}(a)+A_{2}(a)}{2} -\frac{a^{2N-4} \left(A_{2}(a)-A_{1}(a)\right)^{2}}{4 a_{0}^{2N-4}\left(\sqrt{A_{2}(a_{0})}+\sqrt{A_{1}(a_{0})}\right)^2}.
\label{potencial}
\end{equation}
This potential satisfies that $V(a_0)=0$ and, by using Eq. (\ref{CondEstatico}), also that $V'(a_0)=0$ . The second derivative of Eq. (\ref{potencial}) evaluated at the radius $a_0$ gives
\begin{eqnarray}
V''(a_0)&=& -\frac{(N-2)(2N-3) \left(\sqrt{A_{2}(a_{0})}+\sqrt{A_{1}(a_{0}})\right)^{2}}{2a_{0}^2}\nonumber \\  
&&-\frac{(N-2)(2N-5)\left(\sqrt{A_{2}(a_{0})}-\sqrt{A_{1}(a_{0}})\right)^{2}}{2a_{0}^2}-\frac{\left(A_{2}'(a_{0})-A_{1}'(a_{0})\right)^2}{2 \left(\sqrt{A_{2}(a_{0})}+\sqrt{A_{1}(a_{0})}\right)^{2}}\nonumber \\
&&-\frac{2(N-2)\left(A_{2}(a_0)-A_1(a_0)\right)\left(A_{2}'(a_{0})-A_{1}'(a_{0})\right)}{a_{0}\left(\sqrt{A_{2}(a_{0})}+\sqrt{A_{1}(a_{0})}\right)^2}  \nonumber \\
&&+\frac{A_{1}''(a_{0})+A_{2}''(a_{0})}{2}
-\frac{\left(A_{2}(a_0)-A_1(a_0)\right)\left(A_{2}''(a_{0})-A_{1}''(a_{0})\right)}{2\left(\sqrt{A_{1}(a_{0})}+\sqrt{A_{2}(a_{0})}\right)^{2}},
\label{potencial2der}
\end{eqnarray}
which determines that a static configuration is stable under radial perturbations when $V''(a_0)>0$.

\section{Black hole solutions with charge in $F(R)$ gravity}\label{sol-fr}

The action of $F(R)$ gravity coupled to a power law non--linear electrodynamics reads \cite{tsndim} 
\begin{equation}
I =\frac{1}{16\pi }\int d^{n}x\sqrt{-g}\left(R+ f(R) -\alpha \varepsilon |\mathcal{F}| ^{s}\right) ,
\label{Action}
\end{equation}
where $R+ f(R)=F(R)$ is the gravitational Lagrangian, $\mathcal{F} = \mathcal{F}_{\alpha\beta}\mathcal{F}^{\alpha\beta}$ is the Maxwell invariant, and $\mathcal{F}_{\mu \nu }=\partial _{\mu }\mathcal{A}_{\nu }-\partial _{\nu }\mathcal{A}_{\mu }$ is the electromagnetic tensor field in terms of the gauge potential $\mathcal{A}_{\mu }$. The constant $s=N/4$ is adopted in order to have a traceless $T_{\mu\nu}$, $\alpha$ can be taken as $1$ or $-1$ depending on the Maxwell field considered, and $\varepsilon = \mathrm{sign} (\mathcal{F})$.  

For $N\geq 4$, constant scalar curvature $R_0$, and a purely radial electric field $E(r)=F_{tr}=Q/r^2$, the solution to the field equations obtained from Eq. (\ref{Action}) has the form given by Eq. (\ref{metric-N-sphe}), with the following metric function \cite{tsndim} 
\begin{equation}
A(r)=   1- \frac{2 M}{ r^{N-3}}+ \frac{\alpha  2^{N/4} |Q|^{N/2} }{2(1+f'(R_0)) r^{N-2}}-\frac{R_0 r^2}{(N-1) N} 
\label{AmetricNd}
\end{equation}
where  $Q$ is the electric charge and $M$ the mass. It presents a singularity at $r=0$, where Kretschmann scalar diverges.
For $\alpha=1$, there is a critical value of charge $Q_c$; if $|Q|\le Q_c$  the solution presents an event horizon; for $|Q|>Q_c$ has a naked singularity at the origin. The horizons are determined by the real, positive solutions of the equation $A(r)=0$. For $R_0\le 0$ , the event horizon is given by the largest one of them, while for $R_0>0$, the largest one corresponds to the cosmological horizon, and the second largest one to the event horizon.
For $\alpha=-1$, there is no critical value of charge $Q_c$ nor a naked singularity. In the same way as explained above, the event horizon is given by the largest real and positive solution of $A(r)=0$ when $R_0\le 0$. If  $R_0>0$, the event horizon is determined by the second largest one while the largest solution corresponds to the cosmological horizon.

The $\hat{t}\hat{t}$--component of the traceless energy--momentum tensor of the electromagnetic field associated with Eq. (\ref{Action}) can be expressed in the orthonormal coordinates by
\begin{equation}
T_{\hat{t}\hat{t}}=\frac{\alpha}{32\pi}2^{N/4}|Q|^{N/2}(N-2),
\end{equation}
which, for $N\ge3$, is positive if $\alpha >0$ or negative if otherwise. Since wormholes are constructed in the present work, and they allow the presence of exotic matter, the parameter $\alpha$ is left free to analyze its effect. 

\section{Examples}\label{examples} 

Two examples of thin--shell wormholes using the geometry (\ref{metric-N-sphe}), with the metric function given by Eq. (\ref{AmetricNd}), are presented. To do so, a general spacetime $\mathcal{M}$ is constructed by cutting and pasting two regions: $\mathcal{M}_1$ and $\mathcal{M}_2$. The result of this process is a thin layer of matter with a static radius $a_0$ which is always solution of Eq. (\ref{CondEstatico}) and determines the position of the throat by fulfilling the flare--out condition. Its stability under radial perturbations is given by the sign of the second derivative of the effective potential (\ref{potencial2der}). Due to their cumbersome final expressions, the replacement of the metric in these equations is not shown explicitly.

In order to show the results, the most representative graphics have been chosen for Fig. \ref{fig1}, \ref{fig1zoom}, \ref{fig1_aneg}, \ref{fig2}, \ref{fig2zoom}, and \ref{fig2_aneg}. In them, the solid lines represent the stable solutions, the dashed lines the unstable ones, and the meshed regions indicate where the WEC condition is satisfied (and therefore the solution is made of normal matter). The gray regions lack of physical meaning, since they correspond to the removed sections in the construction of the wormhole that were part of the original metrics which contained horizons and singularities.
Because the increase of the dimensionality of the spacetime only offers a change of scale without altering the qualitative behavior of the solutions, the graphical presentation of the results is limited to $N=5$ and $N=4$ (the dimensionality $N=4$ is included for comparison). The construction has been explored by using $\alpha =1$ and $\alpha  =-1$ as well. In order to prevent the presence of ghosts, this work follows the condition of $F'(R)>0$, for further details, Ref. \cite{bronnikov} is suggested. It is worth noting that the horizontal axis in all figures represents an effective charge which incorporates the mass $M$, the dimension $N$, and $F'(R)$. The counting of the solutions is done by fixing the value of the effective charge and analyzing the number of radii $a_0$ corresponding to it.

\subsection{Wormholes symmetric across the throat}

\begin{figure}
\centering
\includegraphics[width=0.8\textwidth]{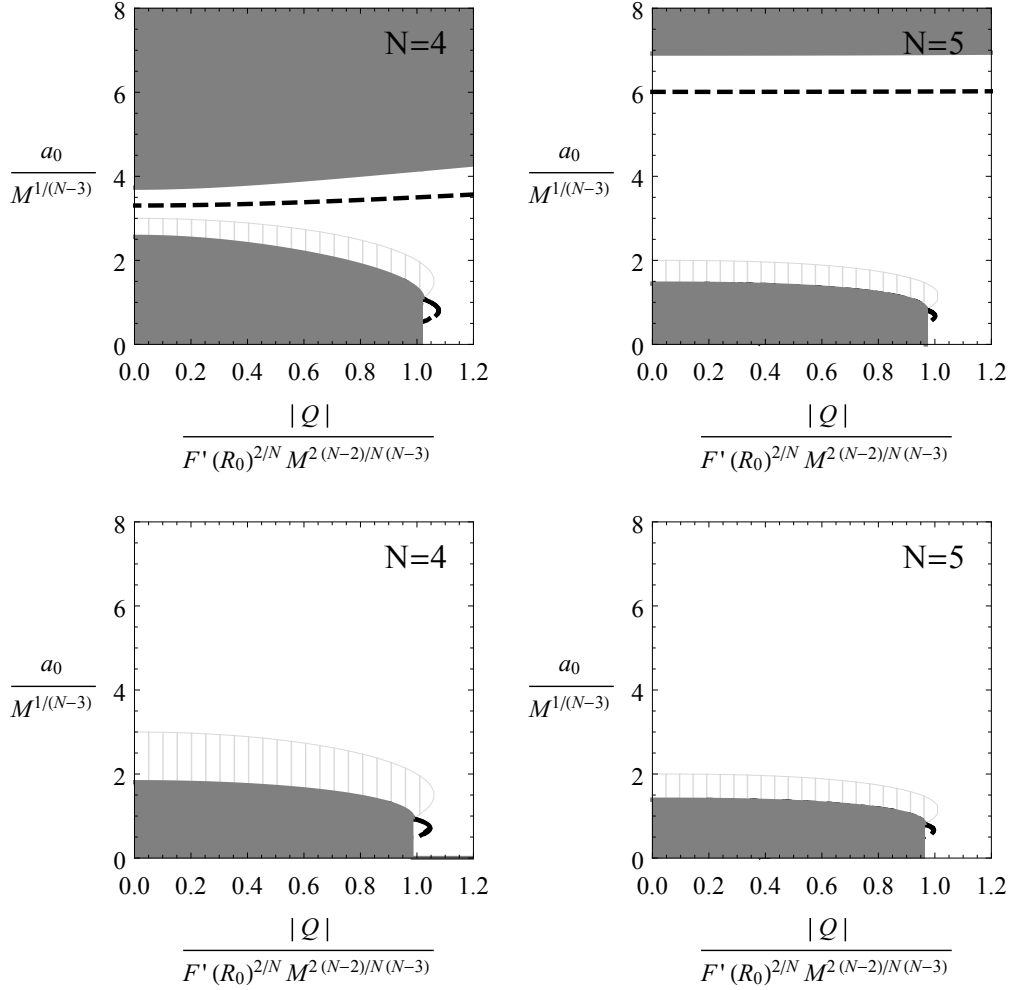}
\caption{Wormholes with $R_1=R_2=R_0$ and $\alpha=1$. The first column shows the results for $N=4$, while the second one for $N=5$. Stable configurations are displayed by solid lines and unstable ones by dashed lines. Meshed regions are related to normal matter, gray zones have no physical meaning. The first row corresponds to plots with $R_0 M^{2/(N-3)}=0.4$ and the second row with $R_0 M^{2/(N-3)}=-0.4$.}
\label{fig1}
\end{figure}

\begin{figure}
\centering
\includegraphics[width=0.4\textwidth]{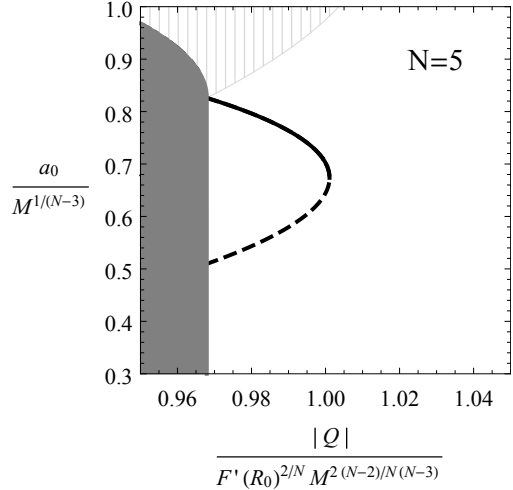}
\caption{Zoom--in of the region around the value $ Q_c$ for wormholes with $\alpha=1$, $N=5$ and $R_0 M^{2/(N-3)}=0.4$ shown in Fig. \ref{fig1}. This behavior is analogous for the rest of the plots displayed in that figure.}
\label{fig1zoom}
\end{figure} 

\begin{figure} [t!]
\centering
\includegraphics[width=0.8\textwidth]{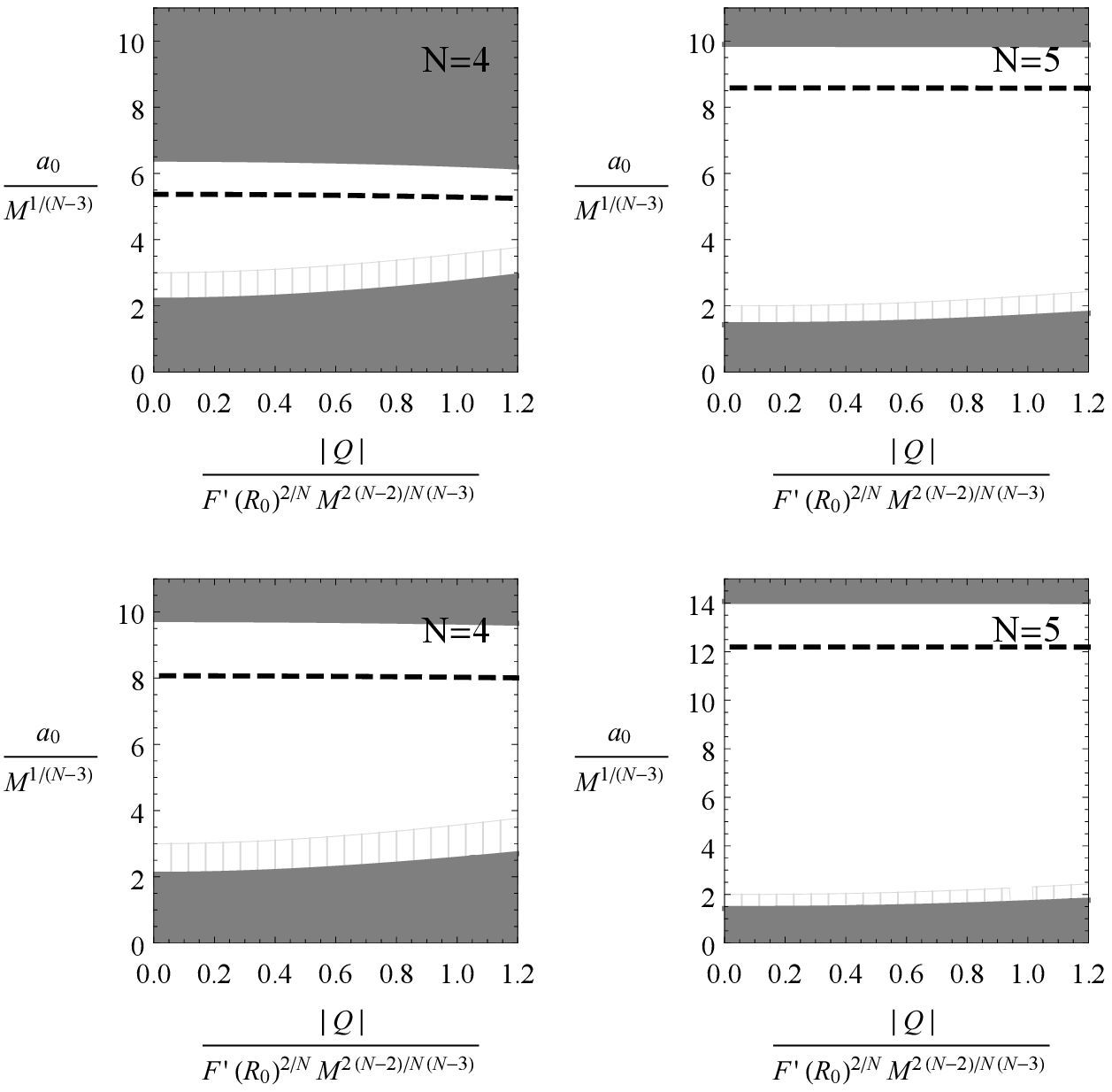}
\caption{Wormholes with $R_1=R_2=R_0$ and $\alpha= -1$. The meaning of $N$, the solid and dashed lines, and the gray and meshed regions are the same as in Fig. \ref{fig1}. The first row corresponds to the plots with $R_0 M^{2/(N-3)}=0.2$ and the second row with $R_0 M^{2/(N-3)}=0.1$.}
\label{fig1_aneg}
\end{figure}

For the construction of this type of wormholes, the metric function given by Eq. (\ref{AmetricNd}) is taken the same at both sides of the throat, this is, $A_1(a)=A_2(a)$, or equivalently, with $M_1=M_2=M$, $Q_1=Q_2=Q$, and $R_1=R_2=R_0$. The static radius of the throat $a_0$ has to be larger than the event horizon and smaller than the cosmological one when it exists. For both values of $\alpha$ there is always a radial electric field  $E(r)=F_{tr}=Q/r^2$.

The results for a wormhole symmetric across the throat with $\alpha = 1$  are shown in Fig. \ref{fig1}, where the columns are related to the dimension of the spacetime, that is, $N=4$ and $N=5$, while the rows correspond to different values of the constant scalar curvature $R_0$. Variations in the mass $M$ of the construction only alter its scale; therefore, all quantities have been adimensionalized with $M$. The meshed regions represent the areas where the WEC condition is fulfilled. For this particular case in Fig. \ref{fig1}, the first row shows the results of $R_0 M^{2/(N-3)}=0.4$ and the second one of $R_0 M^{2/(N-3)}= -0.4$. Due to the difficulty in seeing the behavior of the solutions around  $Q_c$ because the scale of the graphics, Fig. \ref{fig1zoom} has been added. It is a zoom--in of the region around $Q_c$ for the particular case of $R_0 M^{2/(N-3)}=0.4$. The general behavior displayed in it is similar for all the cases presented in Fig. \ref{fig1}.
For  $\alpha = -1$, the results can be seen in Fig. \ref{fig1_aneg}. For this case, only $R_0>0$ has been considered, since there is no solution of Eq. (\ref{CondEstatico}) for $R_0\leq 0$. The figure shows in its first row the results for $R_0 M^{2/(N-3)}=0.2$ and in the second row for $R_0 M^{2/(N-3)}=0.1$. 
The results can be summarized in the following list

\begin{itemize}
\item For  $\alpha = 1$
\begin{itemize}
\item When $R_0>0$ and $|Q| \le Q_c$, there is one unstable solution made of exotic matter close to the cosmological horizon. For $|Q|  \gtrsim  Q_c$, and for a small range of values of charge, there are three solutions made of exotic matter: one of them is stable. For large values of $|Q|$ there is only one unstable solution made of exotic matter.
\item When $R_0\leq 0$ there are only two solutions for a short range of charge $|Q| \gtrsim Q_c$. These solutions are made of exotic matter; the smaller one is unstable while the larger one is stable. 
\item The qualitative behavior of the solutions changes with the sign of the scalar curvature $R_0$, that is, a third solution close to the cosmological horizon may appear depending on this sign.
\item The number of the solutions and their stability behavior do not change with the increase of the dimension $N$. It only produces a change of scale.
\end{itemize}
\item For $\alpha = -1$ 
\begin{itemize}
\item Physical solutions are only possible for $R_0>0$. In this case, only one unstable solution made of exotic matter can be found. It is always close to the cosmological horizon. 
\item The qualitative behavior (only one unstable solution) does not change with the dimensionality $N$, it only produces a change of scale.
\end{itemize}
\item The equation of state for wormholes symmetric across the throat is $\sigma_0 -(N-2)p_0=0$.
\end{itemize}

\subsection{Wormholes asymmetric across the throat}

\begin{figure}
\centering
\includegraphics[width=0.8\textwidth]{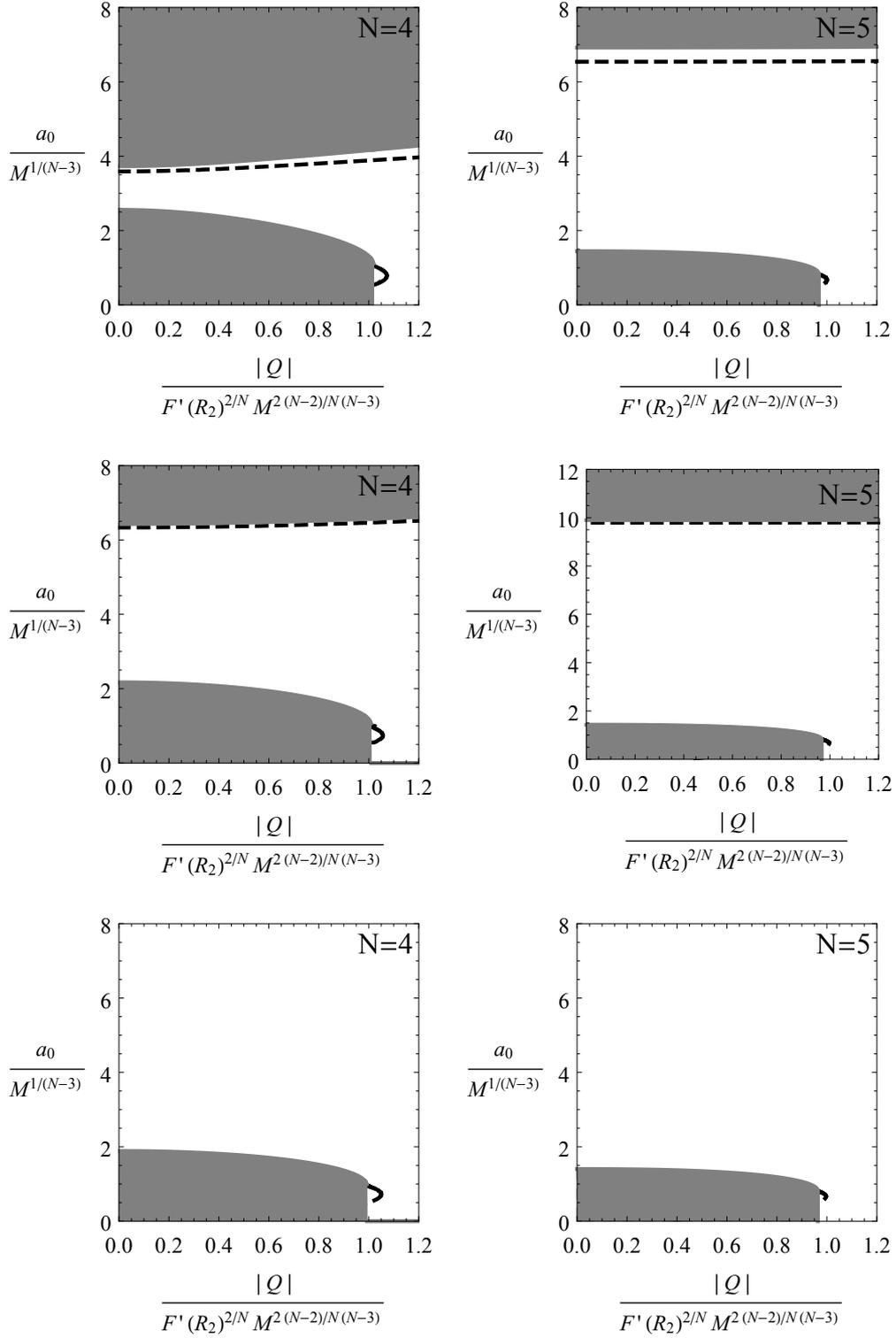}
\caption{Wormholes with $R_1\neq R_2$, $\gamma /M=0.1$, and  $\alpha=1$. The meaning of $N$, the solid and dashed lines, and the gray regions are the same as in Fig.\ref{fig1}. The first row corresponds to $R_1 M^{2/(N-3)}=0.2$ and $R_2 M^{2/(N-3)}=0.4$, the second row to $R_1 M^{2/(N-3)}=0.2$ and $R_2 M^{2/(N-3)}= -0.4$, and the third row to $R_1 M^{2/(N-3)}= -0.2$ and $R_2 M^{2/(N-3)}= -0.4$.}
\label{fig2}
\end{figure}

\begin{figure}
\centering
\includegraphics[width=0.4\textwidth]{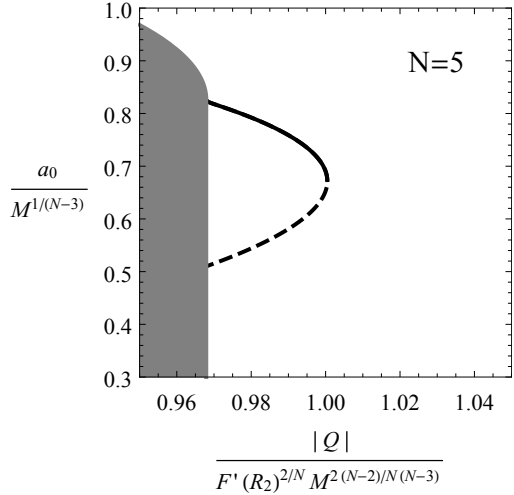}
\caption{Zoom--in of the region around the value $ Q_c$ for wormholes with $\alpha=1$, $\gamma /M=0.1$, $N=5$, $R_1 M^{2/(N-3)}=0.2$ and $R_2 M^{2/(N-3)}=0.4$ shown in Fig. \ref{fig2}. This behavior is analogous for the rest of the cases displayed in that figure.}
\label{fig2zoom}
\end{figure} 

\begin{figure} [t!]
\centering
\includegraphics[width=0.8\textwidth]{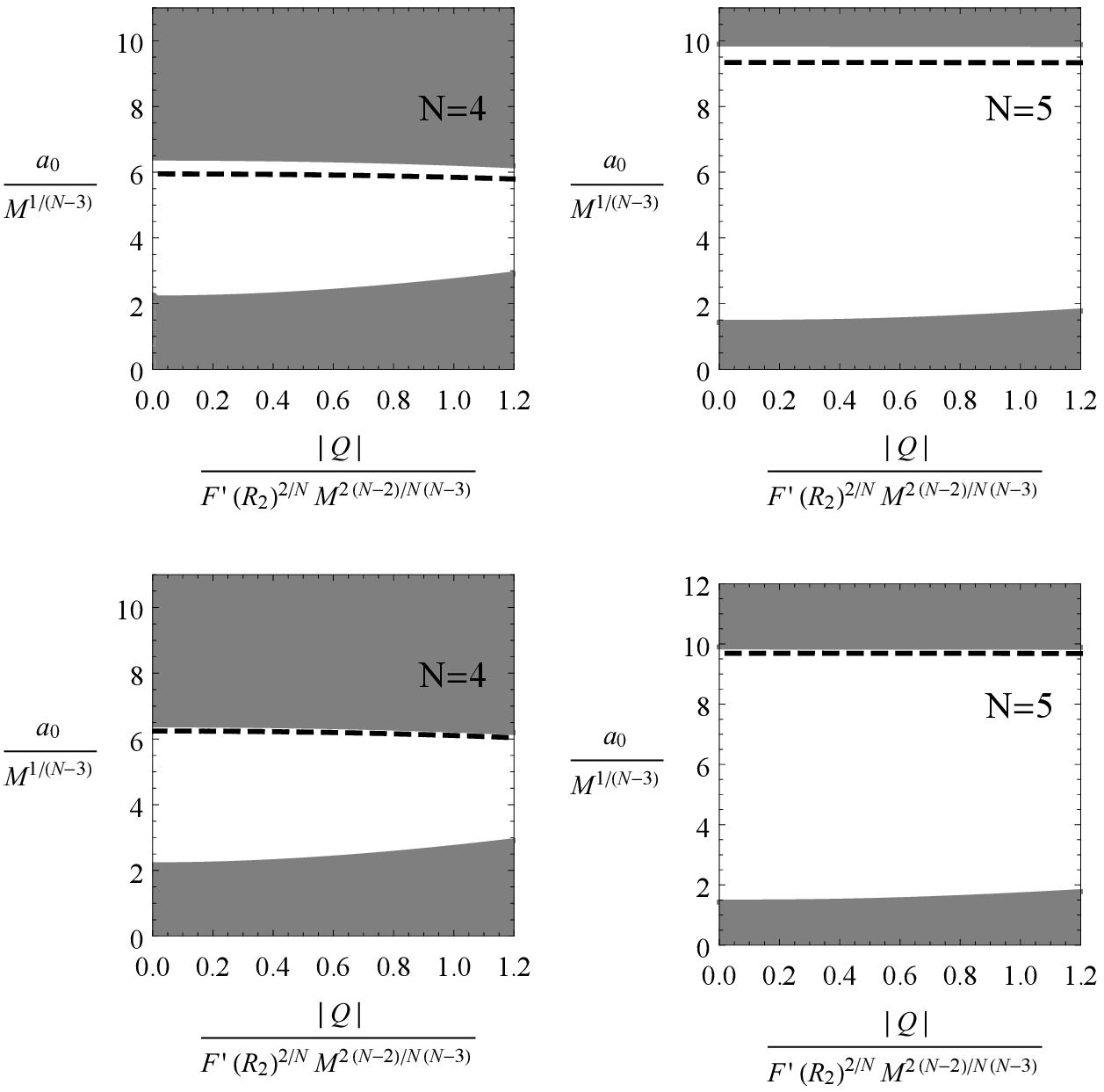}
\caption{Wormholes with $R_1\neq R_2$, $\gamma /M=0.1$, and $\alpha= -1$. The meaning of $N$, the solid and dashed lines, and the gray regions are the same as in Fig. \ref{fig1}. The first row corresponds to $R_1 M^{2/(N-3)}=0.2$ and $R_2 M^{2/(N-3)}=0.1$, and the second row to $R_1 M^{2/(N-3)}=0.2$ and $R_2 M^{2/(N-3)}= -0.1$.}
\label{fig2_aneg}
\end{figure}

For the case asymmetric in the scalar curvature, quadratic theories are required, which allow the relaxation of the continuity of $R$ expressed in Eq. (\ref{3rstcondition}). The function metric for this construction is then given by  Eq. (\ref{AmetricNd}) with $M_1=M_2=M$ and $Q_1=Q_2=Q$, but different constant values of $R_1\neq R_2$ across the throat

\begin{equation}
A_{1,2}(r)= 1-\frac{2 M}{r^{N-3}}+\frac{\alpha 2^{N/4}|Q|^{N/2} }{2(1+2\gamma R_{1,2})r^{N-2} } -\frac{R_{1,2} r^2}{(N-1) N}.
\label{metricasimetric}
\end{equation}
Proceeding in the same fashion as it was done in the previous section, the radius of the throat $a_0$, which is solution of the Eq. (\ref{CondEstatico}), has to be larger than the event horizon and, when $R_{1,2}>0$, smaller than the cosmological one.

The plots show the results for $R_1\neq R_2$; the columns display the dimension, and the rows different combinations of the values of $R_1$ and $R_2$. All quantities have been adimensionalized with $M$, and $\gamma/M= 0.1$ has been taken. The solid and dashed curves, the meshed zones, and the gray regions keep the same interpretation as before.
The graphics in Fig. \ref{fig2} has been done with $\alpha = 1$; the first row corresponds to $R_1 M^{2/(N-3)}=0.2$ and $R_2 M^{2/(N-3)}=0.4$, the second row to $R_1 M^{2/(N-3)}=0.2$ and $R_2 M^{2/(N-3)}= -0.4$, and the third row to $R_1 M^{2/(N-3)}= -0.2$ and $R_2 M^{2/(N-3)}= -0.4$. Because the similar qualitative behavior of the solution around $Q_c$, Fig. \ref{fig2zoom} has been added to better see the details despite the scale. It displays the case for $N=5$, $R_1 M^{2/(N-3)}=0.2$, and $R_2 M^{2/(N-3)}=0.4$. 
The plots in Fig. \ref{fig2_aneg} are for $\alpha = -1$, and present different combinations of $R_{1,2}$ except for $R_{1}<0$ and $R_{2}<0$, in which case there is no physical solution. The columns display the dimensionality of the construction, the first row corresponds to $R_1 M^{2/(N-3)}=0.2$ and $R_2 M^{2/(N-3)}=0.1$, and the second row to $R_1 M^{2/(N-3)}=0.2$ and $R_2 M^{2/(N-3)}= -0.1$. The results can be listed as
\begin{itemize}
\item For  $\alpha = 1$ 
\begin{itemize}
\item When $R_1>0$ or $R_2>0$, and $|Q| \le Q_c$, there is only one unstable solution composed of exotic matter close to the cosmological horizon. For a small range of values of charge $|Q| \gtrsim Q_c$, there are three solutions composed of exotic matter; two of them unstable, and only one stable. For large values of $|Q|$ only one unstable solution made of exotic matter can be found.
\item When $R_1\leq0$ and $R_2\leq0$, there exist two solutions, both made of exotic matter for a short range of charge $|Q| \gtrsim Q_c$. The larger one is stable. 
\item A modification in the dimensionality $N$ of the construction does not change the number of solutions nor their stability behaviour; it only affect its scale. 
\end{itemize}
\item For $\alpha = -1$
\begin{itemize}
\item For $R_1>0$ or $R_2>0$, only one unstable solution made of exotic matter is found. It is close to the cosmological horizon. 
\item For $R_1\leq 0$ and $R_2\leq 0$, physical solutions do not exist. 
\item Again, changes in the dimensionality $N$ only affect the scale.
\end{itemize}
\item For both values of $\alpha$, the external energy flux vector  $\mathcal{T}_\mu$, given by  Eq. (\ref{Tmu}), is zero, and the tensor distribution $\mathcal{T}_{\hat{\imath}\hat{\jmath}}$ is proportional to $2 \gamma (R_2-R_1) h_{\hat{\imath}\hat{\jmath}} /\kappa$. Therefore, the thin shell of matter that constitutes the throat coexists with a double layer.
\item The equation of state for wormholes asymmetric across the throat in quadratic $F(R)$ theories has the form $\sigma - (N-2) p=\mathcal{T}$.
\end{itemize}

\section{Summary}\label{summary}

In this work, a family of wormholes have been built by using a generalized spherical symmetric black hole solution with constant scalar curvature $R$, for $N$--dimensional $F(R)$ gravity in the metric formalism coupled to a conformally invariant Maxwell field. The matter content at the shell has been found. The stability of the static configurations under radial perturbations has been studied, keeping into consideration $F'(R)>0$ to prevent the presence of ghosts. 

Two examples of wormholes have been given, showing the formalism of the construction for $N\geq 4$, a purely radial electromagnetic field with charge $Q$, and a traceless energy--momentum tensor. The two different values of the parameter $\alpha$, that is, $\alpha = 1$ and $\alpha = -1$, have been explored in each of them.
The first example has been constructed within a general $F(R)$, in which case the formalism demands to keep the same value of the constant scalar curvature $R_0$ across the throat. The second example is built within a  quadratic $F(R)$ theory, allowing different values of the constant scalar curvature $R_1 \neq R_2$. In both cases, the expressions of the energy density $\sigma $ and the pressure $p$ at the throat, as well as the equation of state they satisfy have been found. For quadratic $F(R)$, the usual extra contributions that appear in the theory when $R_1 \neq R_2$ have been also shown. In each scenario, the stability of the static configurations for a radius $a_0$ and different combination of the parameters have been analyzed. 

In order to construct a wormhole symmetric across the throat within a general $F(R)$ theory, the same mass, charge, and constant scalar curvature $R_0$ at both sides of the throat have been used. For $\alpha = -1 $ only unstable solutions are found, while for  $\alpha = 1 $, there are stable solutions for a short range of values of the charge. For both values of $\alpha$, they are always made of exotic matter. The behavior of the solutions, that is, the quantity and their stability, depends on the sign of $R_0$ and not on the dimensionality, which only changes their scale.

For the case of wormholes asymmetric across the throat, quadratic $F(R)$ are required, so different constant scalar curvatures $R_1 \neq R_2$ have been used. If both of them $R_{1,2}$ are negative, there are no physical solutions. When $R_1 >0$ or $R_2 >0$, and in similar way to the general $F(R)$ case, for $\alpha = -1 $ only unstable solutions made of exotic matter can be found.   For  $\alpha = 1 $ and a short range of values of the charge, there is a stable solution, which is also made of exotic matter. The behavior of the solutions changes with the relative sign between $R_{1}$ and $R_{2}$, while the dimensionality of the construction $N$ only changes its scale.

\section*{Acknowledgments}

The author thanks to Ernesto F. Eiroa for his useful comments. This work has been supported by CONICET and Universidad de Buenos Aires.

%\section*{Data Availability}
%
%This is a theoretical work in which no experimental data have been used.

\end{document}